\documentclass[runningheads]{llncs}
\usepackage[ruled,vlined]{algorithm2e}
\usepackage[dvipsnames]{xcolor}
\usepackage{graphicx}
\begin{document}

\title{  
A Novel Method for Curating Quanti-Qualitative Content through consensus-based versioning Control
%\thanks{Supported by the National Agency for Research and Innovation (ANII), in Uruguay. MENTOR project FMV-1-2021-1-167914}
}

\titlerunning{A Novel Method for Curating Quanti-Qualitative Content}

\author{Alejandro Adorjan\inst{1}\orcidID{0000-0002-5257-284X} \and
Genoveva Vargas-Solar \inst{2}\orcidID{0000-0001-9545-1821} \and
Regina Motz\inst{3}\orcidID{0000-0002-1426-562X}}
\authorrunning{Adorjan et al.}
\institute{Universidad ORT Uruguay, Montevideo, Uruguay  \\ \email{adorjan@ort.edu.uy}
\and
CNRS Univ Lyon INSA Lyon UCBL, LIRIS UMR5205, Lyon, France
\email{genoveva.vargas-solar@cnrs.fr}\\
 \and
Facultad de Ingeniería,  Universidad de la República,  Montevideo, Uruguay\\
\email{rmotz@fing.edu.uy}}

\maketitle              % typeset the header of the contribution
\begin{abstract}
This paper proposes a Researcher-in-the-Loop (RITL) guided content curation approach for quanti-qualitative research methods that uses a version control system based on consensus.    
The paper introduces a workflow for quanti-qualitative research processes that produces and consumes content versions through collaborative phases validated through consensus protocols performed by research teams.
We argue that content versioning is a critical component that supports the research process's reproducibility, traceability, and rationale. We propose a curation framework that provides methods, protocols and tools for supporting the RITL approach for managing the content produced by quanti-qualitative methods. The paper reports a validation experiment using a use case about the study on disseminating political statements in graffiti.
\end{abstract}
%*********************************************
\section{Introduction}
\label{sec:introduction}
%*********************************************
Quanti-qualitative methods are applied in social and human sciences data-driven research projects. In this context, the production and validation of generated content and knowledge are determined by consensus protocols performed collaboratively by team research \cite{sardana2023qualitative}. Research methods with quanti-qualitative perspectives must curate the content flowing along the research tasks, the processes performed (defining research questions, stating, calibrating and validating a theoretical framework, converging to a corpus, defining data collection tools) and the context and conditions in which content is produced (interviews, focus group, research team decision-making criteria and consensus protocols).

The emergence of data science has led to the use of algorithms to perform specific research tasks like automatically processing surveys using statistics and natural language processing. The conditions in which these automatic processes are used and produce results must be assessed, discussed and consensually validated. Related logs gathering information about the assessment, validation and agreement processes must be curated too. Consider the project willing to classify graffiti images harvested in the downtown area of city to identify those with political content. The objective would be to determine whether anti-system political content can be disseminated as hidden messages in artistic graffiti. In the initial phase of a quanti-qualitative project, a team of scientists would synchronise a specification document with ``rules'' specifying the criteria for choosing people participating in the harvesting process, the area where pictures will be collected and the research question to target. Beyond collecting graffiti pictures, the team would need to maintain the specification documents, information about the people harvesting the data, and the criteria agreed to define the conditions for this harvesting. Moreover, the team would maintain data related to the consensus process that allowed them to decide on the conditions in which graffiti harvesting is done. The consensus process might not be a one-shot task but might need several agreement rounds, during which versions of the specification documents are produced. 

Data curation in the context of the previous example maintains the versions of the content and the comments produced by the research team tagging the documents. The curation result is datasets tracing content about the object of study (i.e., graffiti pictures in a city's downtown) and about the rules agreed upon by a research team to perform graffiti harvesting and information about the consensus process the team followed to make agreements. Curating research artefacts \footnote{We based the notion of artefact according to ACM  definition  ``a digital object that was either created by the authors to be used as part of the study or generated by the experiment itself'', \url{https://www.acm.org/publications/policies/artefact-review-and-badging-current}}.  focusing on the scientific workflow aspects are relevant for reproducibility and improving sciences' reliability \cite{vuorre2018curating}. FAIR (Findable, Accessible, Interoperable and Reusable) principles provide guidelines for scientific data reuse \cite{wilkinson2016fair}. The FAIR guidelines do not fully provide a framework for systematically exploring, curating and reusing research artefacts. Moreover, scientific publication repositories supporting FAIR principles do not answer questions such as:  How do artefacts evolve throughout the research project stages? What software tools and algorithms did researchers rely on? What is the researchers' narrative concerning applying Artificial Intelligence (AI)  algorithms? In the previous example, using AI models for classifying graffiti would require producing explanations on the choice of the algorithms, the principle and the interpretation of their quantitative results, and even the back-and-forth training and calibration phases until the results converge.

%This paper proposes a Version Control System (VCS) for content curation in quanti-qualitative research methods with a Researcher-in-the-Loop (RITL) approach.    
This paper introduces a RITL  workflow for quanti-qualitative research processes. The workflow produces and consumes content versions through collaborative phases validated through consensus protocols performed by research teams.
We argue that content versioning is a critical component that supports the research process's reproducibility, traceability, and rationale. We propose a curation framework that provides methods, practices, and tools for supporting the RITL approach for managing the content produced by quanti-qualitative methods.
%
%This paper proposes a curation approach that considers the production model of content (and knowledge) in data-driven quanti-qualitative projects. 

The contribution of the paper concerns mainly the iterative production and validation of content using consensus protocols performed under RITL settings. Therefore we propose:
\begin{enumerate}
    \item A versioning approach to keep track of the versions of content.
        \item Consensus protocols to control the versioning and track the progress of projects across different stages.
\end{enumerate}

The remainder of the paper is organized as follows. Section \ref{sec:background} introduces the background and related work. Section \ref{sec:overview}    presents a general overview of our proposed content-curation approach for research artefacts produced in quanti-qualitative data-driven research. Section \ref{sec:versioning-control} introduces the RITL-based versioning control model based on consensus protocols that define the management of the curated content. Section \ref{sec:experiments} presents the experimental validation based on two use cases with different content production and validation strategies.
Finally, Section \ref{sec:conclusion} concludes the paper and discusses future work.
%****************************************
\section{Background and Related Work}
\label{sec:background}
%****************************************
Theoretical and methodological frameworks are the basis for sustaining (quali)-qualitative research projects. 
Quantitative methods study observations of phenomena collected  a priori and promote a deductive approach for driving conclusions. For example, analyses numerical data collection methods \cite{sardana2023qualitative}, statistics, machine learning and other artificial intelligence models. 
On the other hand, qualitative approaches  explore complex phenomena. They promote inductive techniques that rely on nonstatistical and nonnumerical data collection methods, analysis, and evidence production \cite{bhangu2023introduction}. 

Quanti-quantitative approaches generally incorporate both methods, addressing complex research questions more accurately due to generalizability, contextualisation, and credibility \cite{sardana2023qualitative}. Quanti-qualitative research methodologies usually apply ``ad-hoc'' data curation strategies that keep track of the data that describe the tools, techniques, hypothesis, and data harvesting criteria. However, these strategies (if present) usually are not standardised nor systematised. In the next lines we give a summarised vision existing data curation techniques and software used to qualitatively curate  content.
%****************************************
\paragraph{Content curation.}
%****************************************
 According to Garcov et al. \cite{garkov2023research}, research data curation is defined as ``the act of  preparing research data and artefacts for sharing and long-term preservation''. Research repositories are the standard for publishing data collections to the research communities. Datasets at an early collection stage are generally not ready for analysis or preservation. Thus, extensive preprocessing, cleaning, transformation, and documentation actions are required to support usability, sharing, and preservation over time \cite{lafia2021leveraging}. We assume that data curation consists of identifying, systematizing,  managing, and versioning research data, considering versioning artefacts an essential component of tracking changes along the research project.
%Anonymous submissions: ICSOC implements a double-blind reviewing process Authors’ prior work should be preferably referred to in the third person; if this is not feasible, the references should be blinded. 
Curated data collections have the  potential to drive scientific progress \cite{zuiderwijk2020drives}. Research artefacts curation is relevant for reproducibility and improves the reliability of sciences \cite{vuorre2018curating}. However, data curation introduces challenges for supporting data-driven applications \cite{esteva2022synchronic} adopting quanti-qualitative methods. For example,  research challenges curating material across time, space and collaborators \cite{vuorre2018curating}. Quantitative and qualitative research methodologies  apply ad-hoc data curation strategies that keep track of the data that describe the tools, techniques, hypothesis, and data harvesting criteria defined a priori by a scientific team. 
%Data sets at an early collection stage are generally not ready for analysis or preservation. Thus, extensive preprocessing, cleaning, transformation, and documentation actions are required to support usability, sharing, and preservation over time \cite{lafia2021leveraging}.
%--------------------------------------
\paragraph{Qualitative Data Processing Software.}
%----------------------------------------
Several software tools that apply statistical techniques and machine learning algorithms are available for qualitative researchers. Woods et al. \cite{woods2016researcher} argue that Computer-Assisted Qualitative Data Analysis Software (CAQDAS) is a well-known tool for qualitative research. These tools support qualitative techniques and methods for applying Qualitative Data Analysis (QDA). ATLAS.ti \cite{atlasti}, Dedoose \cite{dedoose}, MAXQDA \cite{maxqda}, NVivo \cite{nvivo}  implement the REFI-QDA standard, an interoperability exchange format. 
CAQDAS \cite{chen2018using}
researchers and practitioners can perform annotation, labelling, querying, audio and video transcription, pattern discovery, and report generation. Furthermore, CAQDAS tools allow  the creation  of field notes, thematic coding, search for connections, memos (thoughtful comments), contextual analysis, frequency analysis, word location and data analysis presentation in different reporting formats \cite{evers2018current}.  
The REFI-QDA (Rotterdam Exchange Format Initiative) \footnote{https://www.qdasoftware.org}
the standard allows the exchange of qualitative data to enable reuse in QDAS \cite{karcher2021data}. QDA software  such as ATLAS.ti \cite{atlasti}, Dedoose \cite{dedoose}, MAXQDA \cite{maxqda}, NVivo \cite{nvivo}, QDAMiner \cite{qdaminer}, Quirkos \cite{quirkos} and Transana \cite{transana} adopt REFI-QDA standard.

%****************************************
%\paragraph{Researcher-in-the-Loop.}
%****************************************

%RITL is used in various fields, such as data science, machine learning, and artificial intelligence, to ensure that the data used to train models is accurate and unbiased. 
%Several works focus on the Human-in-the-Loop aspect of data science and the improvement of  Artificial Intelligence (AI) models\cite{shang2019democratizing,wang2021putting}.

%. . . . . . . . . . . . . . . . . . . .
\paragraph{Discussion.}
%. . . . . . . . . . . . . . . . . . . .
The researcher's intervention, defined as RITL by \cite{van2020researcher}, is a crucial aspect of human intervention to assess content concerning (i) the conditions in which it is produced and (ii) to make decisions about the new tasks to perform and the way a research project will move forward. Researcher in the loop (RITL) is a case of Human-in-the-loop (HITL), where the primary output of the process is a selection of the data, not a trained machine learning model. HITL is crucial for handling supervision, exception control, optimization, and maintenance \cite{rahwan2018society,mosqueira2023human}. Under a RITL approach, a human sees all data points in the relevant selection at the end of the process. Using RITL requires a systematic solid way of working \footnote{\url{https://hai.stanford.edu/news/humans-loop-design-interactive-ai-systems}}.
This characteristic is critical for designing content curation for quanti-qualitative research methods.

Scientific content should be extracted and computed, including data, analytics tasks (manual and AI models), and associated metadata. This curated content allows the produced knowledge to be reusable and analytics results to be reproducible \cite{leipzig2021role}, thereby adhering to FAIR (Findable, Accessible, Interoperable and Reusable) principles \cite{barcelos2022fair}. 

%However, systems adopting these guidelines are limited to promoting systematic exploring, curating, and reusing research artefacts. Besides, most curation approaches do not collect metadata regarding the protocols adopted by research team members to make decisions and reach a consensus.
% Extensive preprocessing, cleaning, transformation, and documentation actions are required to support usability, sharing, and preservation over time \cite{lafia2021leveraging}. 
%In quanti-qualitative approaches, curation strategies must consider the need to track the different versions of the content produced throughout the research steps. Therefore, it is also required to define version control models to keep track of the changes and evolution of the content \cite{vuorre2018curating,rios2022unifying}. Our work addresses these challenges.
%****************************************
\section{Curating content produced in quanti-qualitative data-driven research}
\label{sec:overview}
%****************************************
We propose a  workflow for modelling (quanti)-qualitative research methods. %The model has five phases: problem statement, data acquisition, data management, data analysis and report. 
%Figure~\ref{fig:design} illustrates the 
It is an spiral $\Gamma$ research process for each phase of a quanti-qualitative research approach: 
% $\Gamma_{1}$ problem statement, $\Gamma_{2}$ data acquisition,$\Gamma_{3}$ data management, $\Gamma_{4}$ data analysis  and $\Gamma_{5}$ report.
% \begin{figure}
% \centering
% \includegraphics[width=0.8\textwidth]{figures/Spiral.png}
% \caption{\label{fig:design} Spiral Qualitative Research Phases.}
% \end{figure}
%\begin{itemize}
    %\item 
    
\noindent 
   - \textit{ $\Gamma_{1}$ Problem statement} refers to the theory review stage, formulation of research questions, definition of methodologies, and construction of the theoretical framework.\\
   % \item 
\noindent 
   - \textit{$\Gamma_{2}$ Data acquisition} is the phase devoted to data collection, exploration, cleaning, and reliability verification. \\
   % \item 
\noindent  
   - \textit{$\Gamma_{3}$ Data management} refers to metadata generation, evaluation, and contextualization.\\
   % \item 
\noindent 
  -  \textit{$\Gamma_{4}$ Analysis} consists of a round of experiments and measurements, incorporating debates and reflections on the results of previous stages. \\
   % \item 
\noindent 
  -  \textit{$\Gamma_{5}$ Reporting} is a phase devoted to visualization, evaluation, writing process, and final publication of scientific work. %\cite{adorjan2022towards}. 
%\end{itemize}

In these phases, research artefacts with content are produced and versioned, guided by consensus-based decision-making by the research team. 
Every phase and its stages adopt a spiral process where artefacts are produced through research actions done within fieldwork activities. For certain activities, research teams agree to use automatic models for extracting meta-data and generating quantitative insight into the content using digital solutions such as algorithms, Mathematical models and text processing tools.

A research team $R$ conducts qualitative activities along these phases and produce content, such as exploratory interviews, weekly meetings reports, manual generation of analytical coding, and analytical coding of the entire process. 
We propose a model representing the content produced during a quati-qualitative research workflow. This content is modelled as artefacts with an associated life cycle: creation, annotation, versioning and validation. A general view of the concepts of the artefact model is given next.
%. . . . . . . . . . . . . . . . . . . .
\subsection{Artefact model}
%. . . . . . . . . . . . . . . . . . . .
Let us define an artefact $\alpha$ as a concept modelling a document like text documents, records, videos or images. 

The artefact $\alpha$ can contain: \\
%\begin{itemize}
    %\item  
\noindent
    - Surveys, interviews, codebooks, field diaries, that are data harvesting ``tools'' used in qualitative methods. \\
   % \item  
\noindent   
   - Descriptions of data harvesting protocols with criteria adopted for choosing focal groups, people applying the tools, objectives, and analysis protocols. \\
    %\item  
\noindent 
   - Documents specifying  theoretical framework, research questions, bibliography, corpora, and quantitative and categorical analysis results. \\
    %\item  
\noindent 
  -  Information describing the context of the (quanti)-qualitative research, including the description of the research team, the conditions in which decision-making is done (agreements, consensus) along the research phases and the provenance of results and agreements. \\
   % \item 
\noindent 
  -  Narratives, multimedia content and technical meta-data (e.g., structure, format, document size) associated with the artefacts.
%\end{itemize}

As stated in the data structure (\ref{eq:artefact}) below, let us denote $\alpha$ as the artefact created at time t represented by a timestamp, in a workflow phase $\Gamma_i$ of a project $\Pi_j$ by a researcher identified with an $Id$. Let $\mu$ represent  
 metadata and $\eta$ represent a narrative annotation (both $\mu$ and $\eta$ are artefacts). An artefact $\alpha$ has associated metadata and narratives.

\begin{equation}
\alpha :
    \begin{tabular}{lcl}
      & & \\
      & $\langle$ & content: Document, \\
      & &producer: ResearcherId, \\
       & &timestamp: Date, \\
       & & projectWfPh: $\langle$ $\Gamma_i$, $\Pi$ $\rangle$ $\rangle$\\
       &  &metaData: [$\mu$], \\
     & &listOfTags: [$\eta$], \\
     & &listOfActions: [$\lambda$] $\rangle$
    \end{tabular}
    \label{eq:artefact}
\end{equation}

%. . . . . .   . . . . . . . .. . .
\noindent
{\em\bf Metadata.} 
%. . . . . .   . . . . . . . .. . .
An artefact $\alpha$ (e.g., a document), by default, brings metadata $\mu$ (title, author, subject, keywords, creation and modification date). We assume that metadata  $\mu$ associated with $\alpha$ artefacts can be extracted automatically and produced by a research team member at a given time through
processing operations $\lambda$ applied to artefacts. For example, an AI algorithms or annotation actions produce metadata that  must be associated with artefacts.

%. . . . . .   . . . . . . . .. . .
\noindent
{\em\bf Narrative.} 
%. . . . . .   . . . . . . . .. . .
A narrative $\eta$ of an artefact $\alpha$ produced by a researcher corresponds to textual content commenting and completing the content of an artefact, or a critical perspective of the content. The narrative is created be a member of the team at a given time $\tau$. 
Let $\eta$ (definition (\ref{eq:narrative-metadata}) below) be an annotation done by a researcher identified with an $Id$ at timestamp t.
\begin{equation}
\noindent \eta :
  \begin{tabular}{l l}

 &  $\langle$ content: Document, \\
 & narrative: Text \\
 &  producer: Researcher, \\
 &  timestamp: $\tau$ $\rangle$\\
 \end{tabular}
 \label{eq:narrative-metadata}
\end{equation}

% Our model represents different types of meta-data $\mu$ to be automatically or manually collected and associated with an artefact.

% \begin{equation}
% \mu :
%    \begin{tabular}{l l}
      
%      & $\langle$ content: Document, \\
%      & ListOfResearchQuestion : Collection (ResearchQuestions), \\
%      & Hypothesis : Collection(Hypothesis), \\
%      & ResearchPhases :  Collection(Phases), \\
%      & ListOfMethods  :  Collection(Methods), \\
%      & TheoreticalFramework : Text, \\
%      & ListOfInstruments : Collection(Instrument), \\
%      & timestamp : Date $\rangle$ \\
%   \end{tabular}
% \end{equation}

%. . . . . . . . . . . . . . . . . ..
\noindent
{\em\bf Action.} 
%. . . . . . . . . . . . . . . . . ..
Let  $\lambda$ represents  the  meta-data of an operation applied on an artefact  (see definition (\ref{eq:algo-metadata}) below). It represents input parameters, chosen by a team member identified with an $Id$, at time $\tau$, with an associated result and a set of assessment scores. Input parameters and results are modelled as artefacts.
\begin{equation}
  \lambda :
   \begin{tabular}{l l}
  &  $\langle$ original: $\alpha$, result: $\alpha$, \\
  &  operation: Operation, \\
 % &  version: Text, \\
  % & url: Text, \\
  &  timestamp: $\tau$ $\rangle$\\
 \end{tabular}
 \label{eq:algo-metadata}
\end{equation}

\begin{table}[h]
    \centering
\begin{tabular}{|l|c|c|p{5cm}|}
  \hline
  \textbf{ID}  & \textbf{Operation} & \textbf{Domain} & \textbf{Description} \\
  \hline
   $Op_{1}$& add ($\alpha$) & $\alpha$,$\sum_{\alpha_{}}^{} \rightarrow \sum_{\alpha_{}}^{}$ & adds $\alpha$ artefact to the  collection of artefacts $\sum_{\alpha_{}}^{} $\\
  \hline
   $Op_{2}$&remove ($\alpha$) & $\alpha$,$\sum_{\alpha_{}}^{} \rightarrow \sum_{\alpha_{}}^{}$   & removes $\alpha$ artefact  from the  collection of artefacts $\sum_{\alpha_{}}^{} $\\
  \hline
   $Op_{3}$&contains($\alpha$) & $\alpha$,$\sum_{\alpha_{}}^{} \rightarrow boolean $ & 
   returns true if the artefact $\alpha$ is in the collection of artefacts $\sum_{\alpha_{}}^{} $\\
  \hline
   $Op_{4}$&addMetadata ($\mu$,$\alpha$)  & 
   $\mu$,$\alpha$ $\rightarrow$$(\mu,\alpha) $&
   adds the metadata $\mu$ to an existing artefact $\alpha$ \\ 
   \hline
   $Op_{5}$&updateMetadata ($\mu$,$\alpha$)  & 
   $\mu$,$\alpha$ $\rightarrow (\mu,\alpha) $&
   updates the metadata $\mu$ of an artefact $\alpha$ \\ 
     \hline
   $Op_{6}$&getMetadata ($\alpha$)  & 
        $\alpha$ $\rightarrow$$(\mu,\alpha) $&
   returns  the metadata $\mu$ of the artefact $\alpha$ \\ 
  \hline
   $Op_{7}$&addRITL ($\eta$,$\lambda$,$\alpha$)  & 
 $\eta$, $\lambda$,$\alpha$ $\rightarrow ((\eta, \lambda),\alpha) $&
   addRITL narrative $\eta$ for the artefact $\alpha$ once applied the algorithm $\lambda$ \\ 
     \hline
   $Op_{8}$&getRITL ($\alpha$)  & 
    $\alpha$ $\rightarrow ((\eta, \lambda),\alpha) $&
   returns the RITL narrative $\eta$ for the artefact $\alpha$ with the corresponding algorithm $\lambda$ \\ 
  \hline
   $Op_{9}$& version ($\sum_{\alpha_{}}^{}$) & $\gamma$,$\sum_{\alpha_{}}^{} \rightarrow \sum_{\alpha_{}}^{}$ & commits  $\gamma$ message to  a new version of the collection of artefacts $\sum_{\alpha_{}}^{} $ denoted as $\sum_{v}$\\
  \hline
   $Op_{10}$& getCurated($url$)  & $ url \rightarrow \sum_{\alpha_{}}^{}$ & given a   $url$  returns a versioned  curated collection of artefacts  $\sum_{v}$\\
  \hline
\end{tabular}
   \caption{Annotation and versioning operations on artefacts.}
    \label{tab:alphaoperators}
\end{table}
Table~\ref{tab:alphaoperators} shows the operations that can be applied to an artefact $\alpha$. An operation is executed at a given time $\tau$ to curate quanti-qualitative content. In general, an operation produces a new version of an artefact timestamped with $\tau$  and signed with the identity of the research team member who executed the operation.
Operations can be classified into three groups: (i) manipulating artefacts adding  and removing an artefact from the collection of a project; (ii) annotating artefacts by adding and updating metadata automatically or by a researcher; (iii) exploring artefacts versions.

%. . . . . . . . . . . . . . . . . ..
\noindent
{\em\bf Project.} 
%. . . . . . . . . . . . . . . . . ..
$\Pi$ is defined as a process consisting of sequential stages of the curation workflow (problem statement, data acquisition, data
management, analysis and report), a collection of artefacts $\sum\nolimits_{\alpha_{}}^{}$,  a research team $R$ and a set of released project versions $\sum\nolimits_{\upsilon_{}}^{}$.

%. . . . . . . . . . . . . . . . . ..
\noindent
{\em\bf Curation process.} 
%. . . . . . . . . . . . . . . . . ..
A version control system implements the artefacts versioning. It traces the life cycle of artefacts and manages artefact versions produced over time by research team members with descriptions of the consensus protocols performed to commit and validate operations. The algorithm \ref{algo:algorithm} shows the curation steps that include creating $\alpha$ artefacts, annotating them with metadata $\mu$ and performing consensus protocols for creating, validating, sharing versions and moving from one workflow step to another.

\begin{algorithm}[H]
    \SetAlgoLined
    \SetKwInOut{Input}{Input}
    \SetKwInOut{Output}{Output}
  
    \Input{: Research artefact $\alpha$}
    \Output{: Version control branch with data, metadata, and narrative in a given collection $\sum_{v}$ from the curated artefact collection $\sum_{\alpha}$}  
    \caption{Curating Research Artefacts}
    \textbf{Step 0}: Initialize the curated collection 
    $\sum_{\alpha} \leftarrow \emptyset$
    
\textbf{Step 1}: Select the research artefact $\alpha$ to be curated and add it to the curated collection$\sum_{\alpha}$
    
    \Indp
     \If{ $\alpha $ $\notin \sum_{\alpha}$}{
        $\sum_{\alpha} \leftarrow \alpha \cup \sum_{\alpha}   $  
        }
    \Indm
    \hspace{2em}\textbf{Step 2}: Create  metadata $\mu$ for the artefact $\alpha$\.
    \Indp
        $\sum_{\alpha} \leftarrow  (\alpha,\mu) $
        adds the corresponding  metadata $\mu$ to the artefact $\alpha$ and to the existing curated collection of artefacts $\sum_{\alpha}$. 
    \Indm   
    
     \hspace{2em}\textbf{Step 3}: 
        \Indp
     Given an operation $\lambda$, an existing research artefact $\alpha$ of the collection $\sum_{\alpha}$, and the narrative $\eta$ produced by a researcher, associate  both the algorithm $\lambda$ and the artefact $\alpha$,
        \hspace{2em} $\sum_{\alpha} \leftarrow   (\alpha,(\lambda, \eta)) $ 
        \Indm
   
    \hspace{2em}\textbf{Step 4: iterate Steps  1, 2 and 3} in a Version Control System.

  \Indp
    \hspace{2em}\textbf{Step 5}: Generate $\sum_{v}$, a curated branch for the artefact collection  $\sum_{\alpha}$.
    \Indm
      \label{algo:algorithm}
\end{algorithm}

The algorithm consists in initialising the containers where the curated artefacts' versions produced during the workflow are gathered and versioned (Step 0). Then in Steps 1 to 4, artefacts' versions are created (Step 1), and meta-data can be associated automatically or manually (Step 2). A subset of artefacts can be then analysed and processed (Step 3), leading to new curated artefacts containing meta-data about the operations (input, output, parameters, results, assessment scores and interpretation narratives). The curation process is performed at every phase of the curation workflow and is iterative until the workflow progresses to the next phase (Step 4). Finally, when a curation iteration converges, a new validated artefact set is liberated and accessible to the whole team (Step 5). 
%***************************************
\section{Consensus-based versioning control} 
\label{sec:versioning-control}
\begin{figure} [htp]
\centering
\includegraphics[width=1\linewidth]{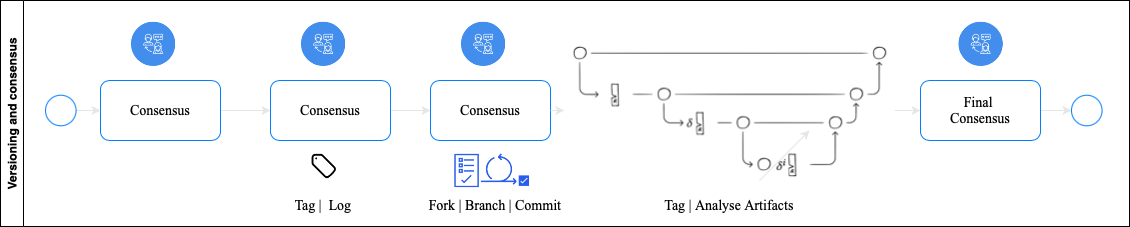}
\caption{Versioning and consensus-based artefacts manipulation.}
\label{fig:versioning}
\end{figure}

%****************************************
A {\em versioning pipeline} and a {\em consensus protocol}  guide the manipulation of artefacts within a project in our approach illustrated in Figure \ref{fig:versioning}. 

The research team chooses the evolution of the status of every stage with its artefacts through a consensus protocol. Through this protocol, the research team members decide whether the status of a project stage will change depending on the validation/disapproval of the artefacts produced in that stage. The research team decides the evolution of the status of every stage of the project and the validation of artefacts through consensus protocols. 
Through these protocols, the research team members decide whether the status of a project stage will change depending on the validation/disapproval of the artefacts $\sum_{\alpha_{}}$ produced in that stage. This version control system implements the artefacts versioning. It traces the life cycle of artefacts and manages artefact versions produced over time by research team members with descriptions of the consensus protocols performed to commit and validate operations. 
 
%---------------------------------------
\subsection{Versioning curated content}
%---------------------------------------
We propose versioning operators to guide a project's workflow for managing curated artefacts $\sum_{\alpha}$ representing content (documents) and metadata.
 Versioning operators represents decisions made at specific project stages to create, validate (globally or by specific team members) and share artefacts' versions $\sum_{v}$. The operators are executed within consensus-based protocols performed by the research team members. This execution model is an original aspect of the versioning control proposed for the quanti-qualitative content curation.  

%...........................................
\noindent { \em Versioning}
%...........................................
We have defined four groups of operators for dealing with the versioning of the stages of a research project. Versioning a stage means versioning the artefacts produced within its different versions as illustrated in Figure \ref{fig:versioning}.

%.......................................
\noindent - {Initialisation $\theta_{init,clone,drop}$.}
%.......................................
The operator $\theta_{init}$  initialises a version workflow with an initial stage with a first branch (main) with a set of artefacts and a history with a first element representing an initial state ( $\theta_{init}$, $\theta_{clone}$).
There are two options for initialising versioning locally $\theta_{init}$  or cloning $\theta_{clone} $ from a remote location.

The operator $\theta_{dropS}$ removes all the tracking versioning of a  workflow leading to the set of artefacts at the stage of the operation $\theta_{dropS}$. The operator  $\theta_{dropB}$ removes the branch from a stage with its artefacts.

%............................................
\noindent - {Adding/removing artefacts from a version pool:$\phi_{add,remove}$, $\phi_{add}$ }
%............................................
 Add and remove artefacts from the  version history of a given project stage branch. 

%.......................................
\noindent - {Commit ${\gamma}$.} %cambio de notación a gamma para no confundir con el timestamp
%....................................... 
The operator commits ${\gamma}$ a stage 
adding an element to that stage's versioning curated collection.

%........................................
\noindent {\em Tagging artefacts $\zeta_{tag}$}
%........................................
$\zeta_{tag}$ represents the operation of associating content with an artefact $\alpha$ by creating tags. 
A tag contains
observations and interpretations (understanding) generated automatically and by a human. Tags are $\alpha$  artefacts containing text, interviews, photos, videos, and data in general. 

%.......................................
\noindent - {Branch ${\beta}$}
%....................................... 
 Forks the main branch of a stage to create a new branch within the versioning workflow. The new branch is associated with a copy of all or part of the artefacts of the stage. It also adds an element to the versioning history of the stage from which the branch is created.

%-----------------------------------
\subsection{Consensus protocols}
%--------------------------------
We proposed in our approach consist of functions \textsc{gpref}(G,i)  and  \textsc{dis}(G, i)  that aggregate individual team members’ preferences to reflect the overall team’s decision-making (according to \cite{amer2019grouptravel} definition on consensus functions for group recommendation)\footnote{Notice that several algorithms could be applied, for example Plurality Voting, Average, Least Misery, Quadratic Voting or Expert Judgement.}. 
In our proposal, for each set of artefacts $\sum_{\alpha_{}}$, of a given research project $\Pi$, the agreement \textsc{gpref}(G,i) or disagreement function \textsc{dis}(G, i) of consensus must be applied for before versioning the curating state of artefacts.

There are two main aspects of a consensus function that refer to group preference and group disagreement: \\
\noindent
{\em (1) Group preference:} The preference of a group for an item needs to reflect the degree to which all group members prefer the item. The more group members prefer an item, the higher its group preference. 
We adopt the average preference  that computes the group preference for an item as the average of individual group members’ preferences for that item \cite{amer2019grouptravel}.  
The preference of an item {\em i} by a group {\em G}, is given by: 
\vspace{-0.2cm}
\begin{equation}
    \textsc{gpref}(G,i) = \frac{1}{|G|}\sum \textsc{pref}(u,i) 
\end{equation}
\noindent
{\em (2) Group disagreement:} The group preference must capture the level at which members disagree or agree.  
 Pair-wise disagreement \cite{amer2019grouptravel}  computes the group preference for an item as the combination of its average and its pair-wise disagreement between individual group members’ preferences. The disagreement of a group $G$ over an item $i$ is given by: 
\vspace{-0.2cm}
\begin{equation}
    \textsc{dis}(G,i) = \frac{2}{|G|(|G|-1)}\sum |\textsc{pref}(u,i)-\textsc{pref}(v,i)| 
\end{equation}
\noindent
 where $u\neq v, u,v \in G$, reflects the degree of consensus in the user-item preferences for $i$ among group members. 
%****************************************
\section{Experimental validation}
\label{sec:experiments}
%****************************************
The first version of the RITL curation approach and framework has been validated in the context of two use cases that apply quanti-qualitative research to perform political and ethnographic data-driven studies.
% project MENTOR (seMantic Exploration aNd curaTion of  Open Hybrid Research)  %supported by the National Agency for Research and Innovation (ANII), in Uruguay, instantiating the following studies :
\footnote{The project is funded by **to be completed if accepted**.}
\begin{enumerate}
    \item $UC_{1}$ Political, artistic project for detecting political statements within grafitti in the downtown of a city in the global south.
 \item  $UC_{2}$ An anthropological study of stud reading practices.
% \item   $UC_{3}$ Analysis of developing digital competencies of higher education abilities in pandemic times. 
%\item   $UC_{2}$ Analysis and understanding of the role of digital environments in constructing gender identities during adolescence in analysing socio-cultural contexts with populations. 
\end{enumerate}
To validate our approach, we first implemented a quanti-qualitative curation framework with tools used by pilot groups of social and human sciences research teams. This paper reports our return of experience on use case $UC_{1}$.
%---------------------------------------
\subsection{Quanti-qualitative curation framework}
\begin{figure}
\centering
\includegraphics[width=\textwidth]{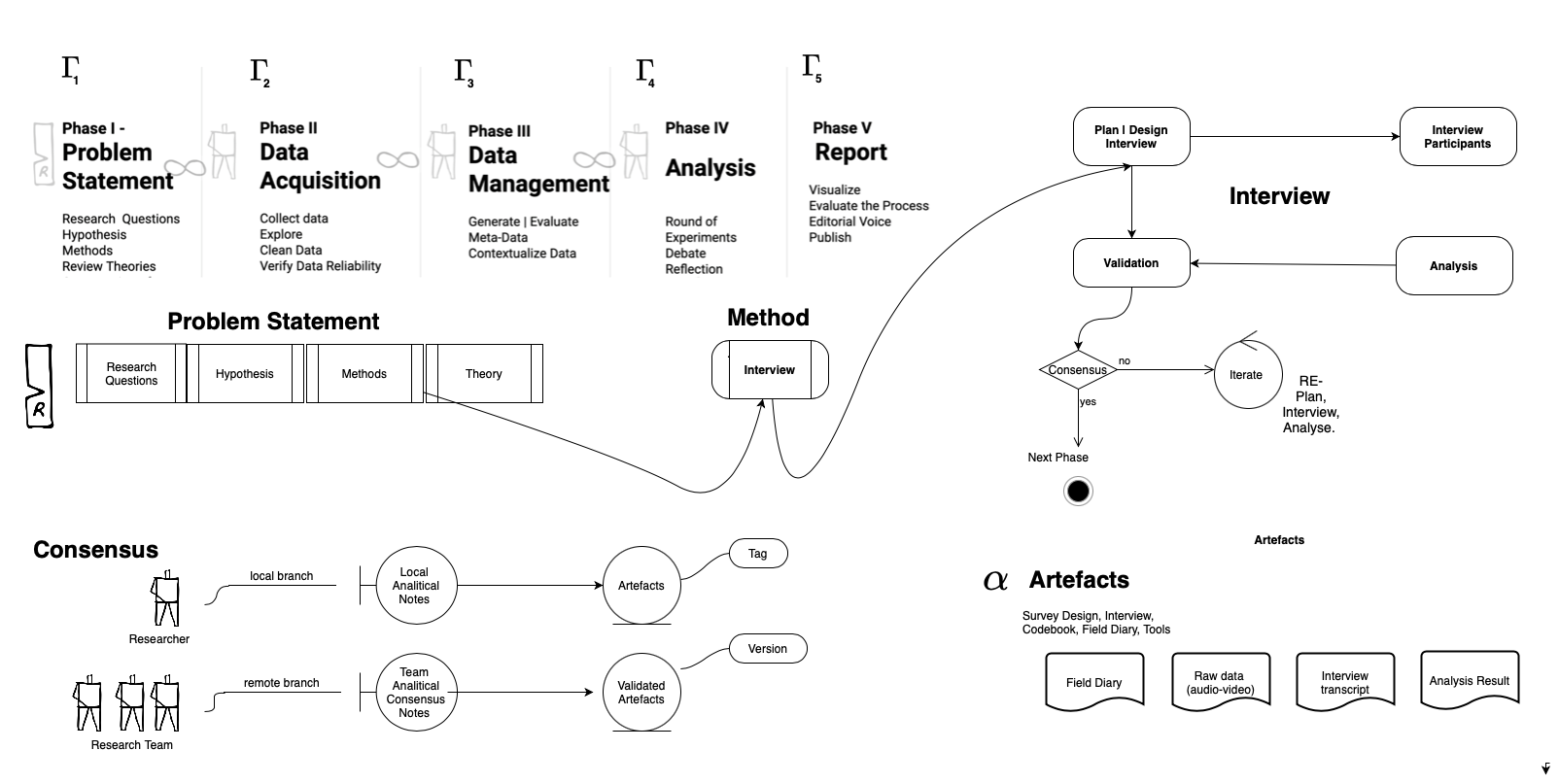}
\caption{RITL driven curation framework for (quanti)-qualitative research methodologies.}
\label{fig:curation-system} 
\end{figure}

%---------------------------------------
We propose a RITL-driven curation system that allows researchers to curate the content produced during the phases of a project (see Figure \ref{fig:curation-system}). It provides tools for uploading digital content, creating associated meta-data (comments, feedback, tags), and interacting asynchronously with other researchers commenting and completing their meta-data. It also provides tools for maintaining the versioning of the content along the different phases of a project and its validation, and also the consensus process that makes the phases evolve from one to another (e.g., branching, validating, committing). The system allows the exploration and analysis of the projects' curated content. Exploration and analysis of curated content will enable them to make informed decisions and have insight into their research practice.

%---------------------------------------
\subsection{Experimental results}
%---------------------------------------
The experimental validation of the framework that implements our content curation approach is intended to observe: (1)  How versioning protocols based on consensus models are integrated into the quanti-qualitative workflows implemented in each use case by research teams. (2) How many cycles performed in each workflow phase took for the research team to converge? (3)  How many artefacts were produced in each use case, and the average number of versions and branches created per phase and cycle of each use case workflow? 
%(4) The type of exploration queries performed for retrieving content versions and the purpose.
According to our notation, $\Gamma$ represents the phases of the research project, proposed by a research team $R$, with a set of curated artefacts  $\sum\nolimits_{\alpha_{}}^{}$,  and a set of released versions  $\sum\nolimits_{\upsilon_{}}^{}$.
Let us define
$\sum_{\eta}$ as  the collection of narratives, regarding a collection of artefacts. Let us represent the research team $R$ implementing the research workflow consisting of team members of different types $R_{i,p}$ with different levels in a hierarchy $p$.
Let us also label the phases of the quanti-qualitative research workflow as the following:  problem statement $\Gamma_{1}$, data acquisition $\Gamma_{2}$, data management $\Gamma_{3}$, data analysis $\Gamma_{4}$  and report $\Gamma_{5}$.

{\bf\em $UC_{1}$ Political content in graffiti:}  The project was held by a research team $R$ with two members, a junior $R_{0}$ and a senior scientist $R_{1}$ working in a hierarchical organisation. The project adopted the spiral qualitative research phases workflow  with the narrative and analytics phases $\Gamma_{3,4}$ merged into one single phase. 

The problem statement phase $\Gamma_{1}$ had two cycles: four versions of an artefact $\alpha$ defining one research question and a narrative containing the definition of political graffiti, inclusion and exclusion rules to select graffiti were enumerated, and the theory framework supporting the pertinence of the research question. In this use case, the final research question emerged from sentence variations: {\em Is it possible to trace and classify political messages disseminated in graffiti in a city?} The first version of the artefact $\alpha_{1}$ had fifteen comments (artefacts $\eta_{1...15}$ proposed by the team members). No branches were created in this phase, and a consensus protocol was implemented until the second cycle using a pairwise disagreement protocol, given the hierarchical organisation and the pair number of members. 

\begin{figure}
\centering
\includegraphics[width=0.8\textwidth]{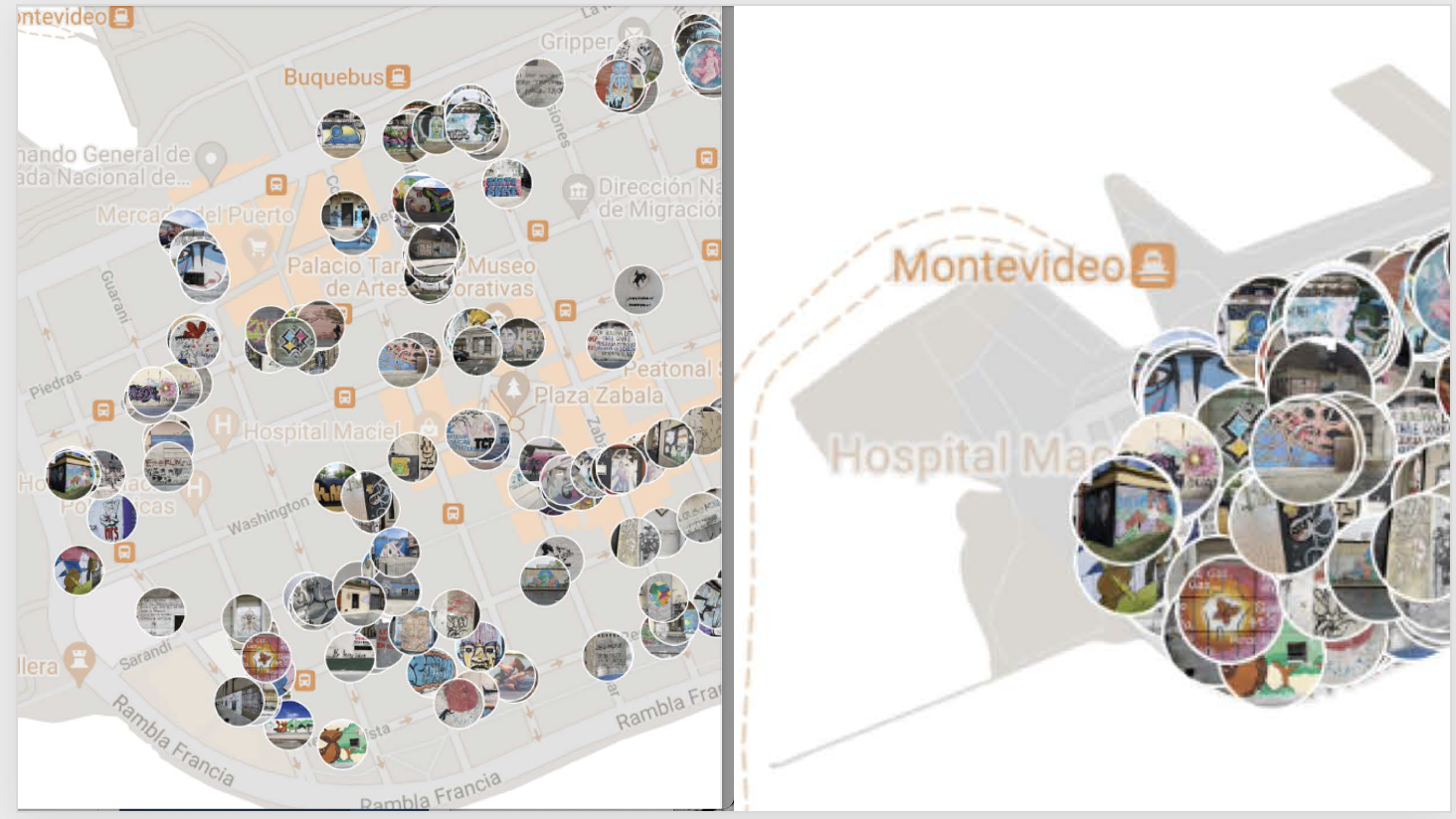}
\label{fig:artefact-2} 
\caption{Graffiti dataset organised according to their geographical area validated in phase $\Gamma_{1}$ graffiti-areas.}
\end{figure}

The data collection phase $\Gamma_{2}$ had three cycles to define the area(s) in the city to collect the photographs of the graffiti by the junior team member. An artefact produced in the first cycle included a set of potential areas in the city to be explored for collecting graffiti samples to acquire evidence to select a data harvesting eventual area. The areas were {\em old town, docks, commercial area, administrative area in downtown}. The sample graffiti sets were artefacts' first versions, with each graffiti tagged with its geographic location. In a second cycle, the team generated artefacts with comments about the elements of the samples. A new consensus pairwise consensus protocol was launched to validate a new artefact with the criteria to select the final collection area. As a result of the consensus, the collection areas were agreed upon in the second cycle. During the third cycle, the junior researcher collected and tagged 1050 graffiti photographs in different downtown areas of the old city. The team validated 546 photos defining a dataset artefact during a consensus pairwise protocol \footnote{The selected final data set is published on Instagram (address to be specified if the paper is accepted).}. An overview of the final dataset organised according to the downtown areas is shown in Figure \ref{fig:artefact-2}. On the right, we can see the graffiti collected downtown, and on the left side, zoom in on those concerning the old city.

The research team developed a phase integrating data management and analysis tasks $\Gamma_{3,4}$. The phase was performed in three cycles. In the first cycle, the team chose an analytics protocol consisting of a classification graffiti task performed in two branches, one with a manual task where an artefact containing classification categories specified how to classify the graffiti validated through a consensus protocol. The second branch consisting in a classification task applying machine learning classification models $\lambda$ provided by Orange \footnote{https://orangedatamining.com}. Therefore the team agreed on k-means and hierarchical classification models ($\lambda$ algorithms). Both branches produced several versions of artefacts classifying the graffiti photographs with associated metadata $\mu$, gathering information about decision-making and assessment scores used to converge into a final classification in each branch.
Metadata $\mu$ containing the comments on the partial results and consensus protocols within the branches were also associated artefacts. In the second cycle, both branches were merged, and the team discussed the classification results generating narratives to interpret them (see Figure \ref{fig:artefact-3}). Four versions of the narratives were produced, including comments and modification requests (i.e., new artefacts) that led to these versions. The narrative with classification was validated through a consensus protocol. The final validated result of the phase $\Gamma_{3,4}$ led to fifteen artefacts. 
\begin{figure}
\centering
\includegraphics[width=0.8\textwidth]{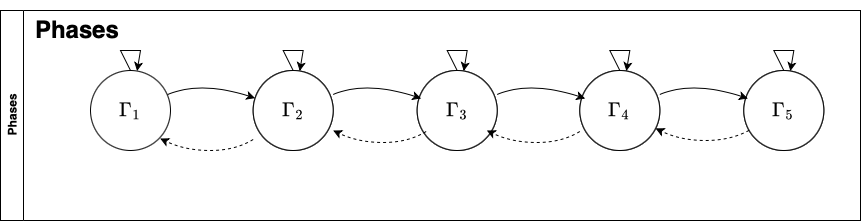}
\label{fig:artefact-3} 
\caption{Phases $\Gamma_{1..5}$.}
\end{figure}

Phases $\Gamma_{1..5}$ evolves also in a versioning cycle begin with $\theta_{init}$ (initialization of the repository), addition and removing of the corresponding artifacts with $\phi_{add,remove}$ operation, and corresponding versioning commit . Subsequently, once the artifacts were tag by $\zeta_{tag}$, the corresponding branches ${\beta}$ were unified to generate the stable version $\sum_{v}$.

Finally, phase $\Gamma_{5}$ consisted of two cycles during which the team produced two versions of the final report with analytics and conclusions about the study. In cycle two, the team validated within a consensus protocol three artefacts, the quantitative results from the classification algorithms, the qualitative results gathered in a report and the graffiti dataset published on Instagram.
\begin{figure}
\centering
\includegraphics[width=0.8\textwidth]{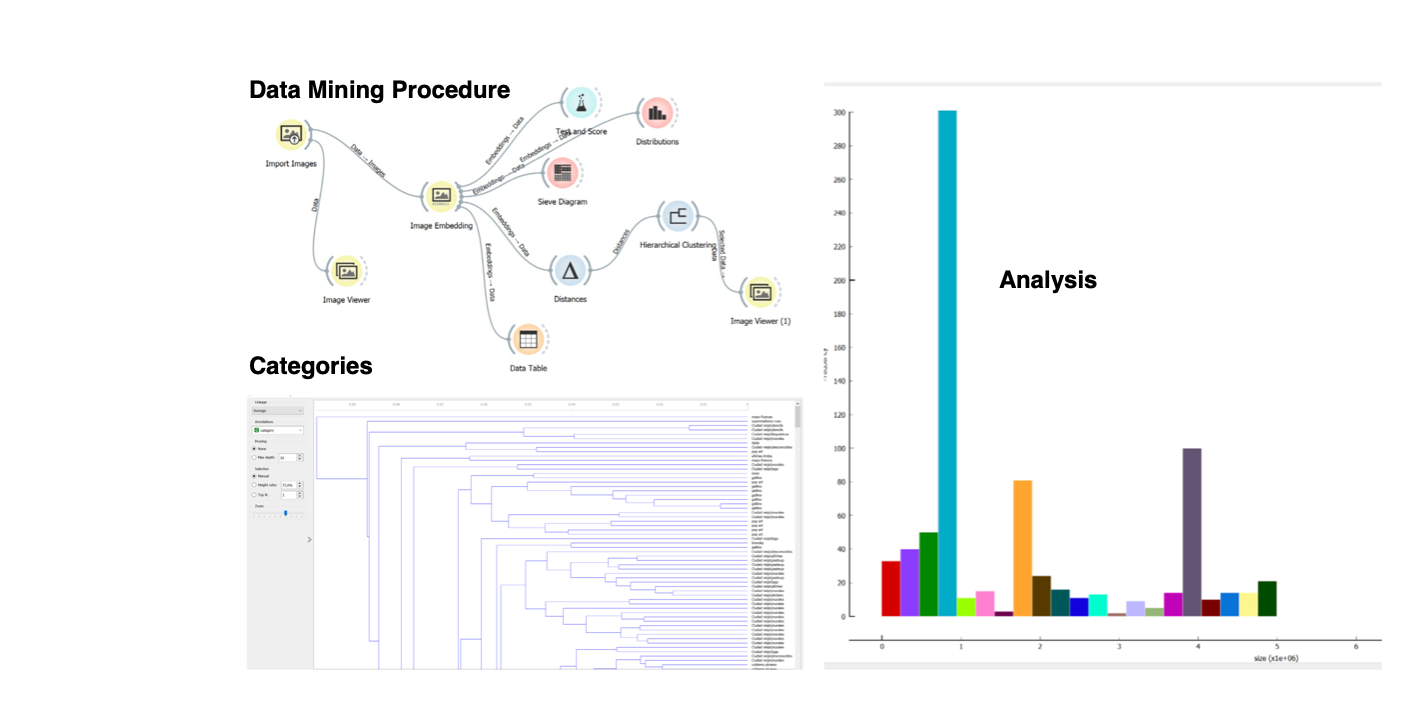}
\label{fig:artefact-3} 
\caption{Analytics artefacts example of the branch 2 of the second cycle of the phase $\Gamma_{3}$ applying $\lambda$ algorithms provided by Data Mining software Orange.}
\end{figure}

\subsection{Lessons learned}
%--------------------------------------
% Table~\ref{tab:stats} shows statistics about the use cases curation. In the first column the Use Case is presented with the corresponding number of artefacts  $\alpha$ of the data collection. The curated collection is represented by $\sum_{\alpha}$, with the corresponding release versions $\sum_{v}$ per phase. The generation of artefacts are also part of each cycle  $\sum $ $\Gamma_{i}$ where the narrative   $\sum_{\eta}$ are proposed by researchers in the senior and junior category. We included the statistics of the use case $UC_{2}$ because it implied creating branches in the problem statement and data collection phases. The use case is implemented by three teams in different regions. Therefore the project workflow was forked. This use case allowed to test these versioning functions.

% \begin{table}
%  \centering   
%   \label{tab:stats}
%   \begin{tabular}{c|c|c|c|c|c|c|c}
%    \hline
%     Use Case & $\alpha$  & Curated $\sum_{\alpha}$ & $\sum_{v}$ per Phase $\Gamma_{1..5}$ & $\sum $ $\Gamma_{1..5}$cycles  & $\sum_{\eta}$  & $[R_{0},R_{1}] $ & \\
%      \hline
%  \em $UC_{1}$ & 1050 & 546 & [4,0,0,4,0] & [2,3,3,3,2] &  15 & [1,1] & \\
%    \em $UC_{2}$ & 60 & 15  & [1,2,1,4,1] & [0,2,4,2,1] &  35 & [3,22] &\\
%    \hline
% \end{tabular}
%  \caption{Curation results for the use cases of the experimental validation.}
% \end{table}
 
In quanti-qualitative data-driven projects, research teams design, application and processing of exploratory interviews for collecting data, define criteria to include and exclude content, annotate content,  and recurrently discuss with the team members to converge on the progress of the project and document discussions and agreements. 

The consensus within the research team allows rigorous analysis and content assessment throughout the process.
Content is contained in artefacts produced as a result of these actions. Besides, these actions can be supported by digital tools such as algorithms, mathematical models, and text processing tools. The criteria used to choose the digital tools and the parameters to calibrate and validate results are also gathered as metadata. 

Researchers' intervention is essential throughout the research process and determines how the activities should be done. Establishing consensus mechanisms as the number of research teams increases is a challenge. 

% Research teams in $UC_{2}$ have asked for lighter ways of performing and curating consensus. Harvesting metadata produced by RITL operations and protocols requires the proposal of semi-automatic solutions. This process is a significant challenge to incorporate into our curation framework.

% When the project needs to be replicated in different countries with the same objective, performing the corresponding fork throughout the project and performing an integrated merge is a good practice.

% In quanti-qualitative research, specifying a single, universal, one-size-fits-all framework that models research requirements is challenging.
% For example, the curation framework in $UC_{2}$ had to be enhanced to manage privacy, security, and legal aspects. However, the solution was technical and should be modelled in the consensus and version management protocols. 

    As the use cases were implemented in our framework, we observed that we must define tools for specifying research workflows according to the characteristics of the projects. The team must have ways to set inputs, outputs of tasks, restrictions, dependencies and other research objectives to guide the curation process. They should also be able to adapt consensus mechanisms for validating artefacts and assessment metrics.

%The RITL-based curation approach promotes the exploration and reproducibility of a qualitative research process by exposing the content production and versioning operations that keep track of the evolution of the artefacts.
%---------------------------------- |
\section{Conclusion and Future Work}
\label{sec:conclusion}
%---------------------------------- |
This paper proposes a  content curation for quanti-qualitative research methods with a RITL approach. The versioning control guides the evolution of a quanti-qualitative research  workflow that produces and consumes content versions through collaborative phases, validated through
consensus protocols performed by the research team. 
The research curation system we propose allows researchers to work on collaborative data-driven projects by processing and managing digital artefacts, their changes, metadata, and narrative throughout the research process. 

We argue that versioning research artefacts with a RITL approach can promote the creation, acquisition, and sharing. This approach also encourages reproducibility, traceability, and rationale of the research process.
Providing a querying model to explore and exploit curated content and workflow phases to derive, reproduce, and trace workflows is part of our future work.
% ---- Bibliography ----
\bibliographystyle{splncs04}
\bibliography{bib}
\end{document}